\documentclass[journal]{IEEEtran}

\ifCLASSINFOpdf

\else

\fi
\usepackage{balance}
\usepackage{lipsum}       
\usepackage{changepage}   
\usepackage{url}
\usepackage{multirow,array}
\usepackage{cite}
\usepackage{graphicx}
\usepackage{hyperref}
\graphicspath{ {Images/} }
\usepackage{caption}
 \usepackage{verbatim}
\usepackage{amsmath}
\usepackage{amssymb}

\usepackage{balance}
\usepackage{lipsum}       
\usepackage{changepage}   
\usepackage{url}
\usepackage{multirow,array}
\usepackage{cite}
\usepackage{graphicx}
\usepackage{hyperref}
\usepackage{caption}
 \usepackage{verbatim}
\usepackage{amsmath}
\usepackage{amssymb}

\usepackage{enumitem}
\usepackage{graphicx}
\usepackage{caption}
\usepackage{subcaption}
\usepackage[numbers,sort&compress]{natbib}
\usepackage{float}

\begin{document}
\title{JPEG Steganalysis Based on DenseNet }

\author{Jianhua Yang,~\IEEEmembership{Student Member,~IEEE,}
        Yun-Qing Shi,~\IEEEmembership{Life Fellow,~IEEE, }\\
         Edward K.Wong,~\IEEEmembership{Senior Member,~IEEE,}
        Xiangui Kang$^*$,~\IEEEmembership{Senior Member,~IEEE}

\thanks{This work was supported by NSFC (Grant Nos. U1536204, 61379155) and
the special funding for basic scientific research of Sun Yat-sen University
(Grant No. 6177060230). (Corresponding author: Xiangui Kang.)

 J. Yang, X. Kang are with Guangdong Key Lab of Information Security, School of Data
and Computer Science, Sun Yat-Sen University, Guangzhou, China 510006, (e-mail:
isskxg@mail.sysu.edu.cn).

Y. Shi is with the department of ECE, New Jersey Institute of Technology, Newark, NJ, USA 07102, (e-mail:shi@njit.edu).

E. Wong is with the department of Computer Science and Engineering, New York University, Tandon School of Engineering, Brooklyn, NY 11201, (e-mail:ewong@nyu.edu).
 }
}


\maketitle

\begin{abstract}

 Different from the conventional deep learning work based on an image¡¯s content in computer vision, deep steganalysis is an art to detect the secret information embedded in an image via deep learning, pose challenge of detection weak information invisible hidden in a host image thus learning in a very low signal-to-noise (SNR) case. In this paper, we propose a 32-layer convolutional neural Networks (CNNs)  in to improve the efficiency of preprocess and reuse the features by concatenating all features from the previous layers with the same feature-map size, thus improve the flow of information and gradient. The shared features and bottleneck layers further improve the feature propagation and reduce the CNN model parameters dramatically. Experimental results on the BOSSbase, BOWS2 and ImageNet datasets have showed that the proposed CNN architecture can improve the performance and enhance the robustness. To further boost the detection accuracy, an ensemble architecture called as CNN-SCA-GFR is proposed,  CNN-SCA-GFR is also the first  work  to combine the CNN architecture and conventional method in the JPEG domain. Experiments show that it can further lower detection errors. Compared with the state-of-the-art method XuNet \cite{XuJPEG2017} on BOSSbase, the proposed CNN-SCA-GFR architecture can reduce detection error rate by 5.67\% for 0.1 bpnzAC and by 4.41\% for 0.4 bpnzAC while the number of training parameters in CNN is only 17\% of what used by XuNet. It also decreases the detection errors from the conventional method SCA-GFR by 7.89\% for 0.1 bpnzAC and 8.06\% for 0.4 bpnzAC, respectively.

\end{abstract}

\begin{IEEEkeywords}
 Adaptive steganography, JPEG steganalysis, convolutional neural Networks (CNNs), ensemble.
\end{IEEEkeywords}


\section{Introduction}
 \setlength{\parskip}{0.1\baselineskip}
 Image steganography is a technology that can hide secret messages in images for covert communication. It includes two research categories: spatial domain and JPEG domain. Current steganography methods for both domains have become more and more sophisticated by
 embedding messages into complex texture regions. In the spatial domain, there are some methods, such as spatial version of UNIWARD (S-UNIWARD) \cite{UNIWARD}, HILL \cite{HILL}, and MiPOD \cite{MiPOD}. Whereas in the JPEG domain, there are some examples, such as UED \cite{UED}, UERD \cite{UERD} and JPEG version of UNIWARD (J-UNIWARD) \cite{UNIWARD}.

 With the development of steganography, steganalysis has also made substantial progress to detect hidden messages in a suspicious image. In the spatial domain, prevailing algorithms calculate residuals from 30 high pass filters, then quantized and truncated the residuals to [-T, T] for co-occurrence operator \cite{SRM, tSRM, maxSRM}. In the JPEG domain, popular algorithms, such as DCTR \cite{DCTR}, GFR \cite{GFR}, PHARM \cite{PHARM} and the relevant versions based on selection-channel-aware \cite{Sca-jpeg} focus on extracting features from the decompressed image. Both in spatial domain and JPEG domain, the ensemble classifiers \cite{Ensemble} is applied for binary classification.

 Recently, image steganalysis has made progress by using convolutional neural networks (CNNs) in the spatial domain \cite{qian2015deep, tan2014stacked,pibre2016deep,sedighi2017histogram,Qian2016, Xustructdesign,Xuensemble,liu2017ensemble,Yang2017, Ni2017}.  In work \cite{Xustructdesign}, Xu \textit{et~al.} find that use Tanh activation to replace ReLU can ensure the features from high pass filters located on the quasi-linear region. In \cite{Ni2017}, Ye \textit{et~al.} used Truncation Linear Layer (TLU) to accelerate the  convergence in spatial steganalysis. However, there are very few works in the JPEG domain \cite{zeng2016large, Chenmo2017, XuJPEG2017}.

 In \cite{zeng2016large}, Zeng \textit{et~al.} proposed a hybrid deep-learning structure based on the large-scale ImageNet dataset. In the preprocessing phase, they used hand-crafted convolutional layers with twenty-five $5\times 5$ DCT patterns from DCTR \cite{DCTR} for decompressed image, then performed quantization and truncation as in conventional steganalysis methods.

 In \cite{Chenmo2017}, Chen \textit{et~al.} proposed a phase-split module to consider the influence of JPEG-phase. To suppresses the image content and increase the high frequency stego signal, four $5\times 5$ high pass filters, which include a ``KV filter'', a ``point filter'', and 2 Gabor filters were used. Experimental results showed that different kinds of high pass filters can complement each other and the detection performance will be improved by using JPEG phase awareness.

 In \cite{XuJPEG2017}, Xu proposed a 20-layer CNN structure based on residual net (ResNet) \cite{ResNet}, he indicated that both the pooling method and network depth are critical for steganalysis. All of the pooling layers were performed by $3\times 3$ convolutional with a stride of 2. To make the CNN architecture deeper, the shortcut connection was incorporated. Experimental results have showed that this method can obtain better results than CNN proposed in \cite{zeng2016large} and conventional method selection-channel aware Gabor filter residuals (SCA-GFR) \cite{Scajpeg}. Thus, we refer to it as XuNet in the rest of this paper.

 In this paper, we study the preprocess and feature reuse for JPEG deep steganalysis. The features can be reused by concatenating all previous layers with matching feature size similiar to the method of Dense Convolutional Network (DenseNet) \cite{DSN}. To further boost the detection accuracy, we propose an ensemble method called CNN-SCA-GFR by combining the CNN architecture and conventional method SCA-GFR.

 The rest of this paper is organized as follows. Details of the proposed CNN architecture are described in Section 2. Experiments and analysis are given in Section 3. Section 4 introduces the proposed ensemble method CNN-SCA-GFR. The conclusion and future works are described in Section 5.

\section{THE PROPOSED JPEG STEGANALYSIS ARCHITECTURE}
The whole architecture are shown on Table \ref{table:CNNSTRUCTURE}. It contain the preprocess layers, features reuse layers and classifier layers. The preprocess layers includes high pass filtering and truncate process.  Features reuse layers were combined by the convolution-batch normalization-ReLU layers. The classifier layers contains a fully-connected layer and a Softmax layer to obtain class probabilities of cover/stego.

\begin {table}[!htb]
\caption {The proposed CNN architecture. Note that each ``conv'' layer shown in the table corresponds to the sequence Conv-BN-ReLU. The size of the convolutional kernels follows the form: (number of kernels $\times$ height $\times$ width).  }
 \label{table:CNNSTRUCTURE}

\begin{center}

\begin{tabular}{ |p{1.2cm}|p{1.4cm}|p{3.50cm}|p{0.6cm}| }

\hline

Group & Output size & Process &Times\\

 \hline
\multirow{1}{*}{Group 1} & \multirow{1}{*}{$256\times 256$}           & High pass filtering&$\times 1$ \\
\hline
\multirow{1}{*}{Group 2} & \multirow{1}{*}{$256\times 256$}           & Truncation&$\times 1$ \\

 \hline
\multirow{2}{*}{Group 3}  & \multirow{2}{*}{$128\times 128$} &       $32\times (3 \times 3)$ conv (stride 1) &\\
                &                           &        $64\times (3 \times 3)$ conv (stride 2) &$\times 1$\\

\hline
\multirow{4}{*}{Group 4} & \multirow{4}{*}{$64\times 64$} & $96\times (1 \times 1)$ conv (stride 1)& \multirow{2}{1cm}{$\times 2$}\\
                                       &    &  $32\times (3 \times 3)$ conv (stride 1)  &   \\ \cline{3-4}

                                       &    &   $128\times (1 \times 1)$ conv (stride 1) &\multirow{2}{0.7cm}{$\times 1$}\\
                                        &   &        $96\times (3 \times 3)$ conv (stride 2) & \\

\hline
\multirow{4}{*}{Group 5} & \multirow{4}{*}{$32\times 32$} & $96\times (1 \times 1)$ conv (stride 1)& \multirow{2}{1cm}{$\times 2$}\\
                                       &    &  $32\times (3 \times 3)$ conv (stride 1)  &   \\ \cline{3-4}

                                       &    &   $128\times (1 \times 1)$ conv (stride 1) &\multirow{2}{0.7cm}{$\times 1$}\\
                                        &   &        $96\times (3 \times 3)$ conv (stride 2) & \\

\hline
\multirow{4}{*}{Group 6} & \multirow{4}{*}{$16\times 16$} & $96\times (1 \times 1)$ conv (stride 1)& \multirow{2}{1cm}{$\times 2$}\\
                                       &    &  $32\times (3 \times 3)$ conv (stride 1)  &   \\ \cline{3-4}

                                       &    &   $128\times (1 \times 1)$ conv (stride 1) &\multirow{2}{0.7cm}{$\times 1$}\\
                                        &   &        $96\times (3 \times 3)$ conv (stride 2) & \\

\hline
\multirow{4}{*}{Group 7} & \multirow{4}{*}{$8\times 8$} &$96\times (1 \times 1)$ conv (stride 1)& \multirow{2}{1cm}{$\times 2$}\\
                                       &    &  $32\times (3 \times 3)$ conv (stride 1)  &   \\ \cline{3-4}

                                       &    &   $128\times (1 \times 1)$ conv (stride 1) &\multirow{2}{0.7cm}{$\times 1$}\\
                                        &   &        $96\times (3 \times 3)$ conv (stride 2) & \\

   \hline
\multirow{4}{*}{Group 8} & \multirow{4}{*}{$1\times 1$} & $96\times (1 \times 1)$ conv (stride 1)& \multirow{2}{1cm}{$\times 2$}\\
                                       &    &  $32\times (3 \times 3)$ conv (stride 1)  &   \\ \cline{3-4}

                                       &    &    $8\times 8$ global average pooling&\multirow{2}{0.7cm}{$\times 1$}\\
                                        &   &         & \\

\hline
\multirow{4}{1cm}{Group 9} &\multirow{4}{1cm}{$1\times 1$} &     \multirow{4}{3cm}{2D fully-connected Softmax}                      & \multirow{4}{0.7cm}{$\times 1$ }\\

                          &                                &  &\\
                                        &   &         & \\
\hline
\end{tabular}
\end{center}

\end {table}

\subsection{High pass filtering}
 In steganalysis, the embedded signal, i.e., hidden information,  is very weak compared with the image content, which is regarded as the embedding noise. That is, it is learning in a very low signal-to-noise (SNR) case. In order to make the CNN architecture concentrate on the weak embedding signals rather than the strong image contents, the JPEG image was decompressed into spatial domain without rounding off the pixel values to integers, then the decompressed image is high-pass filtered via convolution with sixteen high-pass filters which are initialized to DCT basis patterns. The DCT basis patterns are initialized as follows:
   \begin{equation}\label{eq1}
       B_{mn}^{(k,l)}=\frac{w_kw_l}{4}\cos(\frac{k\pi (2m+1)}{8})\cos(\frac{l\pi (2n+1)}{8})
  \end{equation}
 $w_0=1$, $w_x=\frac{1}{\sqrt{2}}$ for $x>0$;$1 \leq k,l \leq 4,1 \leq m,n \leq 4$.\\

After initialization, here in this work, these high-pass filters are optimized through training with other filter parameters rather than fixed \cite{Ni2017}.
\subsection{Truncation Layer}
In this work, the value of the feature maps (x) generated by high pass filters are first truncated with a threshold value of T = 8. TLU \cite{Ni2017} activation function is applied as follows:
\begin{equation}\label{eq2}
f(x)=
\begin{cases}
 -T&   x<-T\\
 x&  -T\leq  x\leqslant T \\
 T&   x>T.
\end{cases}
\end{equation}

 It is observed that applying TLU function can obtain best results in JPEG steganalysis, it is different from the conventional deep learning in computer vision which popularity applies ReLU or Tanh activation function.

\subsection{Feature reuse}
Motivated by the deep learning work of DenseNet \cite{DSN} in conventional computer vision, in this low SNR deep learning work, the features learned from previous layers are also reused by concatenation layers. The feature-maps from all of the preceding layers can be used by all subsequent layers, thus the flow of information and gradients throughout the network are improved.

 Eq. (\ref{eqdese}) illustrates the dense connection manner, where the ${\ell^{th}}$ layer receives the feature-maps of all preceding layers $x_0,...,x_{\ell-1}$ as input:
  \begin{equation}\label{eqdese}
   x_\ell=H_\ell([x_0, x_1, ...,x_{\ell-1}])
  \end{equation}
  where [$x_0,...,x_{\ell-1}$] refers to the concatenation of the feature-maps produced in layers  $0,...,\ell-1$. $H_\ell$ represents the serial process of layer $\ell$ with Convolution, Batch Normalization\cite{BN}, ReLU (Conv-BN-ReLU).

\begin{figure}[!htb]
\centering
\includegraphics[totalheight=4.5in, width=3in, origin=c]{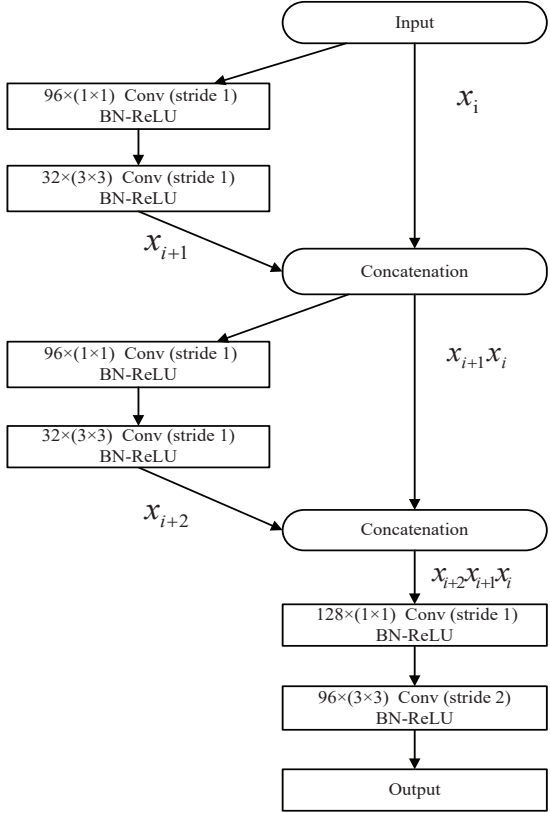}
\caption{The concatenating process of Group 4. }\label{Proposed_group4}
\end{figure}

    Figure. \ref{Proposed_group4} show the Group 4 of the proposed CNN architecture. Please note that we doesn't apply the DenseNet directly, the architecture is varied according to the target of low SNR steganalysis task. All of the average pooling layers were also replaced by convolution layers with stride 2 to propagate the weakly signal. Because of the GPU memory, each convolution layer is followed by a Batch Normalization (BN) and ReLU function (Conv-BN-ReLU), we also doesn't conduct the concatenate process on Group 3.

   Comparing with current best steganalysis method XuNet, the main difference are as follows.

   1) During the preprocess phase, the parameters of sixteen DCT kernels are updated in our method while the parameters are fixed in XuNet.

   2) We ignore the absolution layer after the truncation process.

   3) The $1\times 1$ bottleneck layer and transition layer have been employed to improve computation efficiency and decrease the number of parameters.

   4) The proposed CNN architecture reuses the features by concatenating all previous layers with the same feature-map size. In XuNet, the identity features and the output of non-linear transformation are combined by performing element-wise addition, so that the propagation of information will be impeded and lead to optimization problems \cite{DSN}.

\section{EXPERIMENTS AND ANALYSIS}

\subsection{Platform and Hyperparameters Settings}
 Caffe toolbox \cite{CAFFE} was selected to implement the proposed CNN architecture. Parameters were updated by stochastic gradient descent (SGD). A mini-batch of 32 images with 16 cover-stego pairs were used as the input for each training iterations. The momentum was set to 0.9 and the learning rate was initialized with a value 0.001, then divided by 5 every 30,000 iterations. The convolutional kernels were initialized using a zero-mean Gaussian distribution with a standard deviation 0.01, except that the fully connected layers were initialized using ``Xavier'' initialization. The biases are initialized to 0.2 in the $3\times 3$ convolutions and disabled in the $1\times 1$ convolutions. 

\subsection{Datasets}
\noindent\textbf{BOSSbase}. We use the standard dataset Bossbase v1.01 \cite{BOSSBase} which contains 10,000 cover images. All of the images have been resampled to a size of $256\times 256$ by using ``imresize()'' function in Matlab and then JPEG compressed by using ``imwrite()'' function with a quality factor 75. We selected 6,000 images for training and the remaining 4,000 images were used for validation. All images in the training set were randomly horizontally mirrored and rotated by a multiple of 90 degrees for data augmentation.

 \noindent\textbf{BOWS2}.
  The 10,000  images from dataset BOWS2 \cite{BOWS2} were resampled to size of $256\times 256$ and then JPEG compressed with a quality factor 75 as processed on BOSSbase. All 10,000 images were used for testing the performance of models which trained by dataset BOSSbase.

\noindent\textbf{ImageNet}. In order to verify the performance on large scale dataset, 500,000 images were randomly selected from ImageNet ILSVRC 2013 classification dataset \cite{ImagNet}. We use 80\% images for training, 10\% for validation and 10\% for testing. All the images were cropped from the left-top region of the original image to size $256\times 256$, then convert to grayscale and JPEG recompressed with a quality factor 75.

\begin{figure*}[htbp]
\begin{subfigure}[t]{0.5\textwidth}
\centering
\includegraphics[totalheight=2.5in, width=3.0in, origin=c]{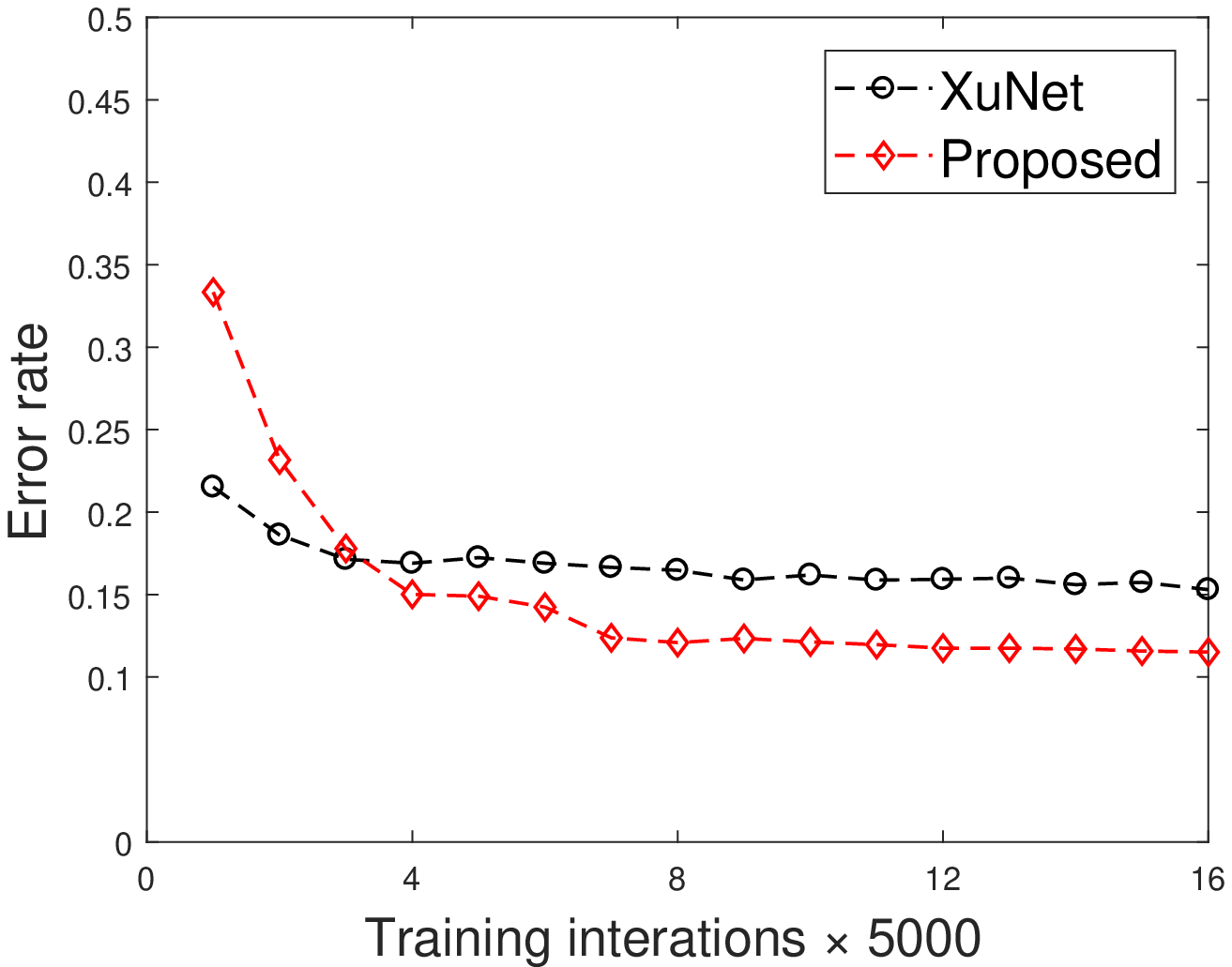}
\caption{}
\label{Xu_proposed_boss}
\end{subfigure}\hspace*{\fill}
\begin{subfigure}[t]{0.5\textwidth}
\centering
\includegraphics[totalheight=2.5in, width=3.0in, origin=c]{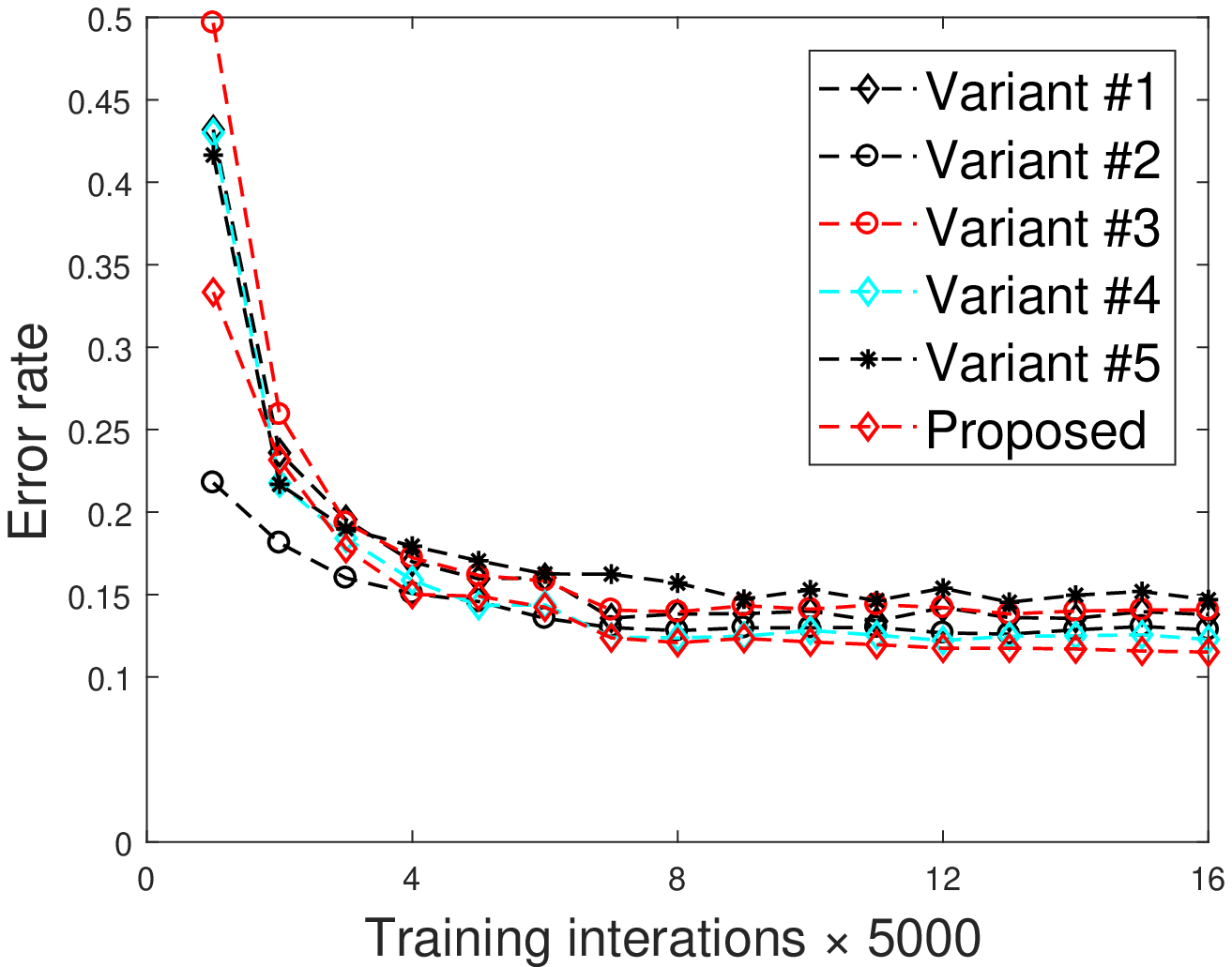}
\caption{}
\label{var1_6_proposed}
\end{subfigure}\hspace*{\fill}

\caption{Detection errors of different methods on Bossbase for J-UNIWARD at 0.4 bpnzAC. (a) the comparison of the proposed method and XuNet, (b) different variants of the proposed method.} \label{fig:variant}
\end{figure*}

 \subsection{ Classification Results on BOSSbase}
 Due to computational constraints in our facilities, we only compare our results with XuNet and the conventional steganalysis methods SCA-GFR for J-UNIWARD with payloads ranging from 0.1 bpnzAC to 0.4 bpnzAC. We trained the CNN architecture 3 times separately with 120,000 iterations, saved the trained models every 5,000 iterations. We select the last 3 saved models from each experiment, and report the ensemble results from 9 trained models to reduce the impact of random variations. The validation error rates from the ensemble results are shown in Table \ref{compare_with_previous}. It can be seen that the proposed architecture can reduce the detection errors dramatically than XuNet and SCA-GFR.
\begin {table}[!ht]
 \caption {Detection error rates (\%) for J-UNIWARD on BOSSbase.   }\label{compare_with_previous}

\begin{center}
\begin{tabular}{ |p{1.4cm}<{\centering}|p{1.4cm}<{\centering}|p{1.8cm}<{\centering}|p{1.8cm}<{\centering}| }
\hline
 Payload & Proposed   &XuNet \cite{XuJPEG2017}  &SCA-GFR \cite{Scajpeg} \\
 \hline
 0.4 bpnzAC &  \textbf{10.84}  &14.16 & 17.81\\
 \hline
0.3 bpnzAC   &\textbf{16.61}     &20.30       &  25.22  \\
 \hline
0.2 bpnzAC & \textbf{25.72} &	29.27 & 33.91 \\
 \hline
0.1 bpnzAC &  \textbf{37.30} & 41.63	 &	43.85  \\

 \hline
\end{tabular}
\end{center}
\end {table}

 In order to investigate the influence caused by different parts of the proposed CNN architecture, we vary the CNN architecture as follows:

Variant \#1: Replace all of the concatenation layer with added layer;

Variant \#2: Delete all of the $96\times (1 \times 1)$ bottleneck layer and $128\times (1 \times 1)$ transition layers;

Variant \#3: Replace all of the $96\times (1 \times 1)$ bottleneck layer and $128\times (1 \times 1)$ transition layers with $96\times (3 \times 3)$ and $128\times (3 \times 3)$ convolution layers;

Variant \#4: Fixed the parameters of the sixteen DCT kernels during the training stage.

Variant \#5: Add an absolute layer after the high pass filters.

The validation errors after 80,000 iterations on BOSSbase for J-UNIWARD at 0.4 bpnzAC are shown in Figure \ref{fig:variant}. The lowest validation errors for comparison have been shown in Table \ref{tab:var}. It can be seen that the proposed method can obtain better performance than XuNet and different variants.


\begin {table}[!ht]
 \caption {Detection error (\%) for different variants on the BOSSbase for J-UNIWARD at 0.4 bpnzAC. }\label{tab:var}

\begin{center}
\begin{tabular}{ |p{0.9cm}|p{0.8cm}|p{0.6cm}|p{0.6cm}|p{0.6cm}|p{0.6cm}|p{0.6cm}| }

\hline
 Proposed  &XuNet  & \# 1  & \# 2    &\# 3  &\# 4      &\# 5   \\

 \hline

\textbf{11.51}    & 15.30 &13.39 & 12.60    &13.83 &12.21   &14.54 \\

\hline

\end{tabular}
\end{center}
\end {table}
\subsection{Analysis}

From variant \#1, replace all of the concatenation layer with added layer will decreased the detection accuracy. This process will make the propagation of information  be impeded and lead to optimization problems [29].

From variant \#2 and \#3, delete all of the $1 \times 1$ bottleneck layer and transition layer or replaced by $3 \times 3$ convolutions will also decreased the detection accuracy, this result may caused by the dramatically increment of training parameters and more harder to train.

From variant \#4, update the DCT kernels during the training stage will improve the detection performance. This process may make the high pass filters more adaptively.


From variant \#5, it's interesting that add an absolute layer will increase the detection accuracy for spatial steganalysis \cite{Xustructdesign}, but the performance will decreased for JPEG steganalysis.

 Because the feature reuse and the utilization of bottleneck layer and transition layer, the architecture become more compactly. Thus the parameter is only 17\% of what used by XuNet (0.88 M versus 5.48 M).

 \subsection{Classification Results on BOWS2}
 In order to investigate  the robustness of the proposed CNN architecture, we test detection errors on the BOWS2 dataset for models trained on the BOSSbase dataset. Experimental results are presented in Table \ref{tab:mismatch}, from which it can be observed that the proposed method is robust for different datasets and can obtain better performance than XuNet and SCA-GFR.

\begin {table}[!ht]
 \caption { Detection error rates (\%) for J-UNIWARD on BOWS2. }\label{tab:mismatch}

\begin{center}
\begin{tabular}{ |p{1.4cm}<{\centering}|p{1.4cm}<{\centering} |p{1.8cm}<{\centering}|p{1.8cm}<{\centering}| }

\hline
 Payload & Proposed   &XuNet \cite{XuJPEG2017} &SCA-GFR \cite{Scajpeg} \\

 \hline
0.4 bpnzAC  & \textbf{15.05}	  &18.41   & 21.05 \\
 \hline
 0.3 bpnzAC  &\textbf{21.70}	& 25.02	 & 28.07  \\
\hline
0.2 bpnzAC  & \textbf{29.91}	& 33.92	  & 35.55 \\
 \hline
0.1 bpnzAC  &\textbf{40.26}	& 44.19	&43.99  \\

\hline
\end{tabular}
\end{center}
\end {table}

 \subsection{Classification Results on ImageNet}
\begin{figure}[!htb]
\centering
\includegraphics[totalheight=2.1in, width=2.5in, origin=c]{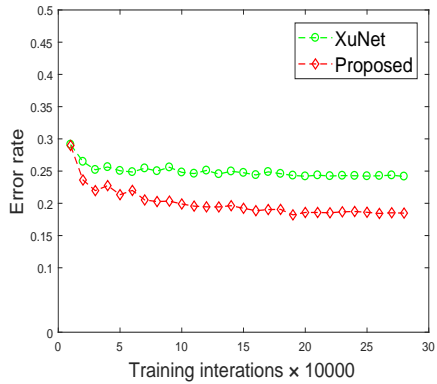}
\caption{Comparison of validation error rate vs training iterations between the proposed CNN and XuNet \cite{XuJPEG2017} on ImageNet for J-UNIWARD at 0.4 bpnzAC.  }\label{Fig110wImagenet}
\end{figure}

We trained models 280,000 iterations for 0.4 bpnzAC, results are shown in Figure \ref{Fig110wImagenet}. The models for 0.2 bpnzAC were fine-tuned from the models optimized from 0.4 bpnzAC \cite{Qian2016}.

Test results based on models with best validation accuracy are shown in Table \ref{tab:imagenet}. It is noted that the performance in terms of detection accuracy will be enhanced on the large scale dataset ImageNet if the batch size can be increased. To be fair and considering the constraint of GPU memory, here in the experiments, we only input 20 cover-stego pairs for both the XuNet and the proposed CNN architecture. It has been observed from Table \ref{tab:imagenet} that the proposed method can also achieve lower error rates on large scale dataset.

\begin {table}[!htb]
\caption {Detection error (\%) compared on ImageNet for J-UNIWARD. }\label{tab:imagenet}
\begin{center}
\begin{tabular}{ |p{2.0cm}|p{2.0cm}<{\centering}|p{2.0cm}<{\centering}|p{2.0cm}<{\centering}| }
\hline
 Algorithm & 0.2 bpnzAC &0.4 bpnzAC     \\
\hline
Proposed & \textbf{33.04} &   \textbf{15.34}  \\
\hline
XuNet\cite{XuJPEG2017}       & 38.68& 22.28     \\
\hline

\end{tabular}
\end{center}
\end {table}

\section{Ensemble CNN with Conventional Method}

\begin{figure*}[t]
\centering
\includegraphics[totalheight=3.5in, width=4.0in, origin=c]{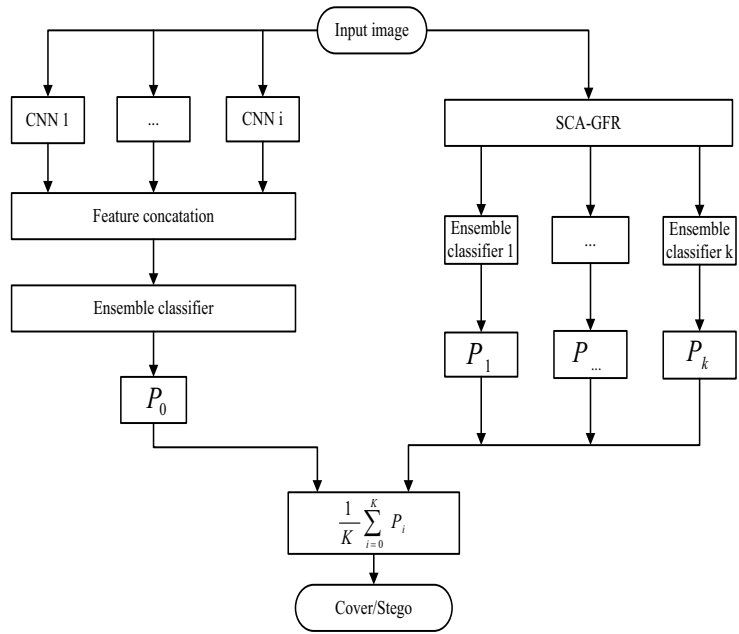}
\caption{Block diagram of the proposed CNN-SCA-GFR architecture. }\label{EnsembleStructure}
\end{figure*}

\begin {table*}[t]
 \caption {Detection error rates (\%) of CNN-SCA-GFR on BOSSbase for J-UNIWARD.}\label{tab:enseble}
\begin{center}
 \begin{tabular}{ |p{2.0cm}<{\centering}|p{2.0cm}<{\centering}|p{1.8cm}<{\centering}|p{1.8cm}<{\centering}|p{1.8cm}<{\centering}| }
\hline
 Payload  &CNN-SCA-GFR  &XuNet \cite{XuJPEG2017}  &SCA-GFR \cite{Scajpeg} & Proposed CNN \\
 \hline

0.4 bpnzAC & \textbf{9.75}     &14.16  &17.81   & 10.84 \\
 \hline

0.3 bpnzAC  &\textbf{15.7}   &20.30    & 25.22   &  16.61 \\
 \hline

0.2 bpnzAC  &\textbf{24.25}   &  29.27  & 33.91 &  25.72\\
 \hline

0.1 bpnzAC  & \textbf{35.96} &41.63   &  43.85 &  37.30 \\
 \hline
\end{tabular}
\end{center}
\end {table*}
Here, we investigate the ensemble approach to combine the proposed CNN architecture with the conventional JPEG steganalysis method SCA-GFR on BOSSbase dataset. The proposed ensemble architecture is shown in Figure \ref{EnsembleStructure}, we set $i = 9$ and $k=6$ by considering the diversity and complexity. For each trained CNN model, 160 dimensional intermediate features are extracted from average pooling layer \cite{Xuensemble}. Then, 1,440 dimensional features from 9 CNNs are concatenated to generate final dimensional features. From SCA-GFR, 17,000 dimensional features are calculated from selection-channel aware variant of GFR feature extractors. Calculate the probability $P_0$ to $P_6$ by using the ensemble classifier. The final class probabilities can be merged by taking the average of 7 probabilities, thus determined the image as cover or stego.

 Different from previous ensemble method in the spatial domain which obtain the probability from Softmax layer \cite{liu2017ensemble} in CNN architecture, here in this work, intermediate features in CNN architecture are used as inputs to the ensemble classifier to obtain probabilities for further process. We call this method CNN-SCA-GFR.

 Experimental results of CNN-SCA-GFR by analysing J-UNIWARD are shown in Table \ref{tab:enseble}. It can be seen that the proposed method can decrease the detection errors from XuNet by 5.67\% for 0.1 bpnzAC and by 4.41\% for 0.4 bpnzAC, respectively. It also decreases the detection errors from the conventional method SCA-GFR by 7.89\% for 0.1 bpnzAC and 8.06\% for 0.4 bpnzAC, respectively.

\section{Conclusion}
In this paper, we have proposed an architecture to detect the JPEG steganalysis signal on low signal-to-noise (SNR) situation. The proposed 32-layer CNN architecture can improve the efficiency of preprocess and reuse the features by concatenating all of the previous layers with the same feature-map size, thus improve the flow of information and decrease the training parameters dramatically. We have also proposed the CNN-SCA-GFR method by combining the proposed CNN architecture and the conventional methods SCA-GFR. Our method is the first time to combine the CNN architecture and conventional method in the JPEG domain. Experiments show that CNN-SCA-GFR can further lower detection errors.

In the future, we will incorporate the selection channel aware (SCA) into the CNN architecture as in conventional method for JPEG steganalysis. We will also improve the memory efficient of the proposed CNN architecture. It's also an interesting work to improve the feature propagation for steganalysis in the spatial domain.

\ifCLASSOPTIONcaptionsoff
  \newpage
\fi

\balance
\bibliographystyle{unsrt}
\bibliography{bare_jrnl}

\begin{thebibliography}{10}

\bibitem{XuJPEG2017}
Guanshuo Xu.
\newblock Deep convolutional neural network to detect j-uniward.
\newblock In {\em 5th ACM Workshop Inf. Hiding Multimedia Secur. (IH\&MMSec)},
  2017.

\bibitem{UNIWARD}
Vojt{\v{e}}ch Holub, Jessica Fridrich, and Tom{\'a}{\v{s}} Denemark.
\newblock Universal distortion function for steganography in an arbitrary
  domain.
\newblock {\em EURASIP Journal on Information Security}, 2014(1):1, 2014.

\bibitem{HILL}
Bin Li, Ming Wang, Jiwu Huang, and Xiaolong Li.
\newblock A new cost function for spatial image steganography.
\newblock In {\em Image Processing (ICIP), 2014 IEEE International Conference
  on}, pages 4206--4210. IEEE, 2014.

\bibitem{MiPOD}
Vahid Sedighi, R{\'e}mi Cogranne, and Jessica Fridrich.
\newblock Content-adaptive steganography by minimizing statistical
  detectability.
\newblock {\em IEEE Transactions on Information Forensics and Security},
  11(2):221--234, 2016.

\bibitem{UED}
Linjie Guo, Jiangqun Ni, and Yun~Qing Shi.
\newblock An efficient jpeg steganographic scheme using uniform embedding.
\newblock In {\em Information Forensics and Security (WIFS), 2012 IEEE
  International Workshop on}, pages 169--174. IEEE, 2012.

\bibitem{UERD}
Linjie Guo, Jiangqun Ni, and Yun~Qing Shi.
\newblock Uniform embedding for efficient jpeg steganography.
\newblock {\em IEEE transactions on Information Forensics and Security},
  9(5):814--825, 2014.

\bibitem{SRM}
Jessica Fridrich and Jan Kodovsky.
\newblock Rich models for steganalysis of digital images.
\newblock {\em IEEE Transactions on Information Forensics and Security},
  7(3):868--882, 2012.

\bibitem{tSRM}
Weixuan Tang, Haodong Li, Weiqi Luo, and Jiwu Huang.
\newblock Adaptive steganalysis against wow embedding algorithm.
\newblock In {\em Proceedings of the 2nd ACM workshop on Information hiding and
  multimedia security}, pages 91--96. ACM, 2014.

\bibitem{maxSRM}
Tomas Denemark, Vahid Sedighi, Vojtech Holub, R{\'e}mi Cogranne, and Jessica
  Fridrich.
\newblock Selection-channel-aware rich model for steganalysis of digital
  images.
\newblock In {\em Information Forensics and Security (WIFS), 2014 IEEE
  International Workshop on}, pages 48--53. IEEE, 2014.

\bibitem{DCTR}
V.~Holub and J.~Fridrich.
\newblock Low-complexity features for jpeg steganalysis using undecimated dct.
\newblock {\em IEEE Transactions on Information Forensics and Security},
  10(2):219--228, Feb 2015.

\bibitem{GFR}
Xiaofeng Song, Fenlin Liu, Chunfang Yang, Xiangyang Luo, and Yi~Zhang.
\newblock Steganalysis of adaptive jpeg steganography using 2d gabor filters.
\newblock In {\em Proceedings of the 3rd ACM Workshop on Information Hiding and
  Multimedia Security}, pages 15--23. ACM, 2015.

\bibitem{PHARM}
Vojtech Holub and Jessica~J Fridrich.
\newblock Phase-aware projection model for steganalysis of jpeg images.
\newblock In {\em Media Watermarking, Security, and Forensics}, page 94090T,
  2015.

\bibitem{Sca-jpeg}
Tom{\'a}{\v{s}}~Denemark Denemark, Mehdi Boroumand, and Jessica Fridrich.
\newblock Steganalysis features for content-adaptive jpeg steganography.
\newblock {\em IEEE Transactions on Information Forensics and Security},
  11(8):1736--1746, 2016.

\bibitem{Ensemble}
Jan Kodovsky, Jessica Fridrich, and Vojt{\v{e}}ch Holub.
\newblock Ensemble classifiers for steganalysis of digital media.
\newblock {\em IEEE Transactions on Information Forensics and Security},
  7(2):432--444, 2012.

\bibitem{qian2015deep}
Yinlong Qian, Jing Dong, Wei Wang, and Tieniu Tan.
\newblock Deep learning for steganalysis via convolutional neural networks.
\newblock {\em Media Watermarking, Security, and Forensics},
  9409:94090J--94090J, 2015.

\bibitem{tan2014stacked}
Shunquan Tan and Bin Li.
\newblock Stacked convolutional auto-encoders for steganalysis of digital
  images.
\newblock In {\em Asia-Pacific Signal and Information Processing Association,
  2014 Annual Summit and Conference (APSIPA)}, pages 1--4. IEEE, 2014.

\bibitem{pibre2016deep}
Lionel Pibre, J{\'e}r{\^o}me Pasquet, Dino Ienco, and Marc Chaumont.
\newblock Deep learning is a good steganalysis tool when embedding key is
  reused for different images, even if there is a cover source mismatch.
\newblock {\em Electronic Imaging}, 2016(8):1--11, 2016.

\bibitem{sedighi2017histogram}
Vahid Sedighi and Jessica Fridrich.
\newblock Histogram layer, moving convolutional neural networks towards
  feature-based steganalysis.
\newblock {\em Electronic Imaging}, 2017(7):50--55, 2017.

\bibitem{Qian2016}
Yinlong Qian, Jing Dong, Wei Wang, and Tieniu Tan.
\newblock Learning and transferring representations for image steganalysis
  using convolutional neural network.
\newblock In {\em Image Processing (ICIP), 2016 IEEE International Conference
  on}, pages 2752--2756. IEEE, 2016.

\bibitem{Xustructdesign}
Guanshuo Xu, Han-Zhou Wu, and Yun-Qing Shi.
\newblock Structural design of convolutional neural networks for steganalysis.
\newblock {\em IEEE Signal Processing Letters}, 23(5):708--712, 2016.

\bibitem{Xuensemble}
Guanshuo Xu, Han-Zhou Wu, and Yun-Qing~Shi ~.
\newblock Ensemble of cnns for steganalysis: an empirical study.
\newblock In {\em Proceedings of the 4th ACM Workshop on Information Hiding and
  Multimedia Security}, pages 103--107. ACM, 2016.

\bibitem{liu2017ensemble}
Kai Liu, Jianhua Yang, and Xiangui Kang.
\newblock Ensemble of cnn and rich model for steganalysis.
\newblock In {\em Systems, Signals and Image Processing (IWSSIP), 2017
  International Conference on}, pages 1--5. IEEE, 2017.

\bibitem{Yang2017}
Jianhua Yang, Kai Liu, Xiangui Kang, Edward Wong, and Yunqing Shi.
\newblock Steganalysis based on awareness of selection-channel and deep
  learning.
\newblock In {\em International Workshop on Digital Watermarking}, pages
  263--272. Springer, 2017.

\bibitem{Ni2017}
Jian Ye, Jiangqun Ni, and Yang Yi.
\newblock Deep learning hierarchical representations for image steganalysis.
\newblock {\em IEEE Transactions on Information Forensics and Security},
  12(11):2545--2557, 2017.

\bibitem{zeng2016large}
Jishen Zeng, Shunquan Tan, Bin Li, and Jiwu Huang.
\newblock Large-scale jpeg steganalysis using hybrid deep-learning framework.
\newblock {\em arXiv preprint arXiv:1611.03233}, 2016.

\bibitem{Chenmo2017}
Mo~Chen, Vahid Sedighi, Mehdi Boroumand, and Jessica Fridrich.
\newblock Jpeg-phase-aware convolutional neural network for steganalysis of
  jpeg images.
\newblock In {\em 5th ACM Workshop Inf. Hiding Multimedia Secur. (IH\&MMSec)},
  2017.

\bibitem{ResNet}
Kaiming He, Xiangyu Zhang, Shaoqing Ren, and Jian Sun.
\newblock Deep residual learning for image recognition.
\newblock In {\em Proceedings of the IEEE conference on computer vision and
  pattern recognition}, pages 770--778, 2016.

\bibitem{Scajpeg}
T.~D. Denemark, M.~Boroumand, and J.~Fridrich.
\newblock Steganalysis features for content-adaptive jpeg steganography.
\newblock {\em IEEE Transactions on Information Forensics and Security},
  11(8):1736--1746, Aug 2016.

\bibitem{DSN}
Gao Huang, Zhuang Liu, Laurens van~der Maaten, and Kilian~Q Weinberger.
\newblock Densely connected convolutional networks.
\newblock In {\em Proceedings of the IEEE Conference on Computer Vision and
  Pattern Recognition}, 2017.

\bibitem{BN}
Sergey Ioffe and Christian Szegedy.
\newblock Batch normalization: Accelerating deep network training by reducing
  internal covariate shift.
\newblock In {\em International Conference on Machine Learning}, pages
  448--456, 2015.

\bibitem{CAFFE}
Yangqing Jia, Evan Shelhamer, Jeff Donahue, Sergey Karayev, Jonathan Long, Ross
  Girshick, Sergio Guadarrama, and Trevor Darrell.
\newblock Caffe: Convolutional architecture for fast feature embedding.
\newblock In {\em Proceedings of the 22nd ACM international conference on
  Multimedia}, pages 675--678. ACM, 2014.

\bibitem{BOSSBase}
Patrick Bas, Tom{\'a}{\v{s}} Filler, and Tom{\'a}{\v{s}} Pevn{\`y}.
\newblock " break our steganographic system": The ins and outs of organizing
  boss.
\newblock In {\em Information Hiding}, pages 59--70. Springer, 2011.

\bibitem{BOWS2}
P.~Bas and T.~Furon.
\newblock Bows-2.
\newblock In {\em http://bows2.ec-lille.fr.}, 2007.

\bibitem{ImagNet}
Olga Russakovsky, Jia Deng, Hao Su, Jonathan Krause, Sanjeev Satheesh, Sean Ma,
  Zhiheng Huang, Andrej Karpathy, Aditya Khosla, Michael Bernstein, et~al.
\newblock Imagenet large scale visual recognition challenge.
\newblock {\em International Journal of Computer Vision}, 115(3):211--252,
  2015.

\bibitem{pleiss2017memory}
Geoff Pleiss, Danlu Chen, Gao Huang, Tongcheng Li, Laurens van~der Maaten, and
  Kilian~Q Weinberger.
\newblock Memory-efficient implementation of densenets.
\newblock {\em arXiv preprint arXiv:1707.06990}, 2017.

\end{thebibliography}

\end{document}